# OPTICAL MODES IN PT-SYMMETRIC DOUBLE-CHANNEL WAVEGUIDES

Li CHEN, Rujiang LI, Na YANG, Da CHEN, Lu LI*

Institute of Theoretical Physics, Shanxi University, Taiyuan 030006, China
E-mail: llz@sxu.edu.cn

We investigate the unique properties of various analytical optical modes, including the fundamental modes and the excited modes, in a double-channel waveguide with parity-time (PT) symmetry. Based on these optical modes, the dependence of the threshold values for the gain/loss parameter, i.e., PT symmetry breaking points, on the structure parameters is discussed. We find that the threshold value for the excited modes is larger than that of the fundamental mode. In addition, the beam dynamics in the double-channel waveguide with PT symmetry is also investigated.

## 1. INTRODUCTION

The theoretical and experimental studies on quantum-optical analogies have seen a spectacular resurgence, in which the propagation of optical waves in waveguides and optical lattices has become an important platform to investigate these unique phenomena [1]. For example, in quantum mechanics, one of the fundamental axioms is the Hermiticity of the Hamiltonian operator associated with the physical observable, which not only implies real eigenvalues but also guarantees probability conservation. However, replacing the Hermiticity condition, a new concept of parity-time (PT) symmetry has been proposed in the framework of quantum mechanics [2], and can exhibit entirely real eigenvalue spectra even for the non-Hermitian Hamiltonians [2–5]. Although the impact of PT symmetry in quantum mechanics is still debated, it has been shown that, in optics, PT-related notions can be implemented in PT symmetric coupler [6, 7] and PT symmetric optical lattices [8-10], and the experimental observations have been demonstrated [11–13]. Thus, optics can provide a fertile ground to investigate PT-related beam dynamics including the non-reciprocal responses, the power oscillations, and the optical transparency. Furthermore, PT-related beam dynamics in the nonlinear regimes has been studied extensively, such as, optical solitons in PT-symmetric potentials [8, 14, 15] and unidirectional invisibility induced by PT-symmetric periodic structures [16, 17].

Among these investigations, the dynamics of light-beam propagation in a coupler composed of a double-channel waveguide with PT symmetry gained particular attention because it can exhibit some universal properties in linear and nonlinear regimes [6, 7, 11, 12]. Indeed, the in past years, the double-channel waveguide structure as an important platform has been studied extensively, and the results have shown that the coupler can support symmetry-preserving solutions, which have linear counterparts, and symmetry-breaking solutions without any linear counterparts [18–22]. In this paper, we mainly focus on various analytical stationary solutions in a double-channel waveguide with PT symmetry, and discuss the properties of the symmetry-preserving solutions. In addition, we will also discuss the beam dynamics in the double-channel waveguide with PT symmetry.

The paper is organized as follows. In Sec. II, the model equations describing beam propagation in a double-channel coupled waveguide with PT symmetry are presented. The properties of symmetry-preserving optical modes and the dependence of the PT symmetry breaking points on the structure parameters are discussed in Sec. III. Meanwhile, the beam dynamics in the double-channel waveguide with PT symmetry is also demonstrated. The conclusions are summarized in Sec. IV.

## 2. MODEL EQUATIONS

We consider a planar double-channel waveguide with the complex refractive index distribution $n(x)$, which is of the form $n(x) = n_r - i n_i$ as $-L_0/2 - D_0 < x < -L_0/2$ and $n(x) = n_r + i n_i$ as $L_0/2 < x < L_0/2 + D_0$, otherwise, $n(x) = n_0 (< n_r)$, where $D_0$ and $L_0$ represent the width of channel and the separation between channels, respectively; while $n_0$ and $n_r$ are the substrate index and channel index, respectively, and $n_i$ is the gain/loss parameter. Without loss of generality, here we assume that $n_i > 0$. One can easily see that the complex refractive index distribution $n(x)$ satisfies the necessary condition $n(x) = n^*(-x)$ for PT symmetry. Thus the complex refractive index distribution presents a double-channel waveguide structure with PT symmetry [6,7]. Under slowly varying envelope approximation, the wave equation governing beam propagation in such a waveguide can be written as

$$i \frac{\partial \psi}{\partial z} + \frac{1}{2k} \frac{\partial^2 \psi}{\partial x^2} + \frac{k[n(x) - n_0]}{n_0} \psi = 0, \qquad (1)$$

where $\psi(z, x)$ is the envelope function and $k = 2\pi n_0/\lambda$ is wave number, with $\lambda$ being wavelength of the optical source generating the beam. Equation (1) describes the propagation of an optical beam in a double-channel waveguide with PT symmetry. It should be pointed out that, in contrast to conventional optical systems, the actual total power $P(z) = \int_{-\infty}^{+\infty} |\psi|^2 \, dx$ is no longer conserved in PT symmetric structures, but the "quasi-power" $Q(z) = \int_{-\infty}^{+\infty} \psi(z,x) \psi^*(z,-x) dx$ is a constant of motion independent of distance $z$ [6–10].

Introducing the following normalized transformations $\psi(z, x) = \sqrt{P_0 / l}\, \varphi(\zeta, \xi)$, $\xi = x/l$ and $\zeta = z/2kl^2$, the dimensionless form of Eq. (1) is of the form

$$i \frac{\partial \varphi}{\partial \zeta} + \frac{\partial^2 \varphi}{\partial \xi^2} + V(\xi) \varphi = 0, \qquad (2)$$

Here $V(\xi) = U(\xi) + iW(\xi)$, which describes the dimensionless PT symmetric double-channel waveguide structure, with

$$U(\xi) = \begin{cases} U_0, & L/2 < |\xi| < L/2 + D, \\ 0, & \text{others,} \end{cases} \qquad (3)$$

and

$$W(\xi) = \begin{cases} W_0, & L/2 < \xi < L/2 + D, \\ -W_0, & -L/2 - D < \xi < -L/2, \\ 0, & \text{others,} \end{cases} \qquad (4)$$

where $U_0 = 2k^2 l^2 (n_r - n_0)/n_0$ is the modulation depth of the refractive index, $W_0 = 2k^2 l^2 n_i/n_0$ is the dimensionless gain/loss parameter, and $L = L_0/l$ and $D = D_0/l$ correspond to the scaled separation and width of channels, respectively. Thus, the model (2) presents a dimensionless form for the beam propagation in the PT symmetric double-channel waveguide, in which the refractive index profiles of the double-channel waveguide structure given by Eqs. (3) and (4) are plotted in Fig. 1.

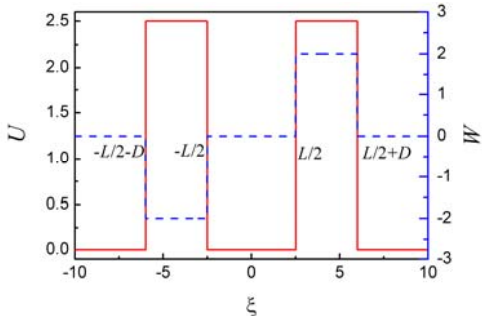

Fig. 1 – The real part (red solid) and the imaginary part (blue dashed) of the refractive index profile of a double-channel waveguide structure with PT symmetry.

We consider the stationary solution for Eq. (2) in the form $\varphi(\zeta, \xi) = \phi(\xi) \exp(i\beta\zeta) \equiv [u(\xi) + v(\xi)] \exp(i\beta\zeta)$ where $u(\xi)$ and $v(\xi)$ are the real and the imaginary parts of $\varphi(\zeta, \xi)$, respectively, and β is the propagation constant. Substituting it into Eq. (2), we can find that the real functions $u(\xi)$ and $v(\xi)$ satisfy the following ordinary differential equation

$$\frac{d^2 u}{d\xi^2} + [U(\xi) - \beta]u = W(\xi)v, \qquad (5)$$

$$\frac{d^2 v}{d\xi^2} + [U(\xi) - \beta]v = -W(\xi)u, \qquad (6)$$

Equations (5) and (6) are a coupled system of differential equations, which is coupled by the gain/loss distributed parameter $W(\xi)$. Here we require that the modes must be normalized to one, i. e. $\int_{-\infty}^{+\infty} |\varphi|^2 \, d\xi = 1$, to ensure their uniqueness for a fixed propagation constant. Thus, we can obtain the optical field modes in double-channel waveguide with PT symmetry by solving Eqs. (5) and (6). In the following Section, we will determine the optical modes and we will discuss their salient properties.

### 3. OPTICAL MODES AND THEIR PROPERTIES

In this Section, we first present the exact solutions of Eqs. (5) and (6). Indeed, in the region of $L/2 < \xi < L/2 + D$, the solutions for Eqs. (5) and (6) are of the form

$$\begin{cases} u_1(\xi; A_1, A_2, \delta_1, \delta_2) = A_1 \exp(\lambda_1 \xi) \sin(\lambda_2 \xi + \delta_1) + A_2 \exp(-\lambda_1 \xi) \sin(\lambda_2 \xi + \delta_2), \\ v_1(\xi; A_1, A_2, \delta_1, \delta_2) = A_1 \exp(\lambda_1 \xi) \cos(\lambda_2 \xi + \delta_1) - A_2 \exp(-\lambda_1 \xi) \cos(\lambda_2 \xi + \delta_2), \end{cases} \qquad (7)$$

where $\lambda_1 = \{[(U_0 - \beta)^2 + W_0^2]^{1/2} - (U_0 - \beta)\}^{1/2}/\sqrt{2}$ and $\lambda_2 = \{[(U_0 - \beta)^2 + W_0^2]^{1/2} + (U_0 - \beta)\}^{1/2}/\sqrt{2}$, and here we have used the assumption $W_0 > 0$ due to $n_i > 0$. In the region $-L/2 - D < \xi < -L/2$, the solutions for Eqs. (5) and (6) are of the form

$$\begin{cases} u_2(\xi; B_1, B_2, \sigma_1, \sigma_2) = B_1 \exp(\lambda_1 \xi) \sin(\lambda_2 \xi + \sigma_1) + B_2 \exp(-\lambda_1 \xi) \sin(\lambda_2 \xi + \sigma_2), \\ v_2(\xi; B_1, B_2, \sigma_1, \sigma_2) = -B_1 \exp(\lambda_1 \xi) \cos(\lambda_2 \xi + \sigma_1) + B_2 \exp(-\lambda_1 \xi) \cos(\lambda_2 \xi + \sigma_2). \end{cases} \qquad (8)$$

In the region $|\xi| < L/2$, the solutions for Eqs. (5) and (6) should be of the form

$$\begin{cases} u_3(\xi; C_{11}, C_{12}) = C_{11} \cosh(\sqrt{\beta}\xi) + C_{12} \sinh(\sqrt{\beta}\xi), \\ v_3(\xi; C_{21}, C_{22}) = C_{21} \cosh(\sqrt{\beta}\xi) + C_{22} \sinh(\sqrt{\beta}\xi). \end{cases} \qquad (9)$$

Also, it is easy to show that the solutions of Eqs. (5) and (6) can be written as

$$\begin{cases} u_4(\xi; D_1) = D_1 e^{-\sqrt{\beta}\xi}, \\ v_4(\xi; D_2) = D_2 e^{-\sqrt{\beta}\xi}, \end{cases} \qquad (10)$$

in the region $\xi > L/2 + D$, and

$$\begin{cases} u_5(\xi; E_1) = E_1 e^{\sqrt{\beta}\xi}, \\ v_5(\xi; E_2) = E_2 e^{\sqrt{\beta}\xi}, \end{cases} \qquad (11)$$

in the region $\xi < -L/2 - D$.

Thus one can construct the analytical global solutions of Eqs. (5) and (6) by employing Eqs. (7–11)

$$u(\xi) = \begin{cases} u_5(\xi; E_1), & \xi < -L/2 - D, \\ u_2(\xi; B_1, B_2, \sigma_1, \sigma_2), & -L/2 - D < \xi < -L/2, \\ u_3(\xi; C_{11}, C_{12}), & |\xi| < L/2, \\ u_1(\xi; A_1, A_2, \delta_1, \delta_2), & L/2 < \xi < L/2 + D, \\ u_4(\xi; D_1), & \xi > L/2 + D, \end{cases} \quad (12)$$

$$v(\xi) = \begin{cases} v_5(\xi; E_2), & \xi < -L/2 - D, \\ v_2(\xi; B_1, B_2, \sigma_1, \sigma_2), & -L/2 - D < \xi < -L/2, \\ v_3(\xi; C_{21}, C_{22}), & |\xi| < L/2, \\ v_1(\xi; A_1, A_2, \delta_1, \delta_2), & L/2 < \xi < L/2 + D, \\ v_4(\xi; D_2), & \xi > L/2 + D, \end{cases}$$

where the continuity conditions of $u$, $v$, $\partial u/\partial \xi$ and $\partial v/\partial \xi$ at the boundaries require

$$\begin{aligned} u_2(-L/2-D; B_1, B_2, \sigma_1, \sigma_2) &= u_5(-L/2-D; E_1), \\ \frac{\mathrm{d}}{\mathrm{d}\xi} u_2(-L/2-D; B_1, B_2, \sigma_1, \sigma_2) &= \frac{\mathrm{d}}{\mathrm{d}\xi} u_5(-L/2-D; E_1), \\ u_2(-L/2; B_1, B_2, \sigma_1, \sigma_2) &= u_3(-L/2; C_{11}, C_{12}), \\ \frac{\mathrm{d}}{\mathrm{d}\xi} u_2(-L/2; B_1, B_2, \sigma_1, \sigma_2) &= \frac{\mathrm{d}}{\mathrm{d}\xi} u_3(-L/2; C_{11}, C_{12}), \\ u_1(L/2; A_1, A_2, \delta_1, \delta_2) &= u_3(L/2; C_{11}, C_{12}), \\ \frac{\mathrm{d}}{\mathrm{d}\xi} u_1(L/2; A_1, A_2, \delta_1, \delta_2) &= \frac{\mathrm{d}}{\mathrm{d}\xi} u_3(L/2; C_{11}, C_{12}), \\ u_1(L/2+D; A_1, A_2, \delta_1, \delta_2) &= u_4(L/2+D; D_1), \\ \frac{\mathrm{d}}{\mathrm{d}\xi} u_1(L/2+D; A_1, A_2, \delta_1, \delta_2) &= \frac{\mathrm{d}}{\mathrm{d}\xi} u_4(L/2+D; D_1), \end{aligned} \quad (13)$$

and

$$\begin{aligned} v_2(-L/2-D; B_1, B_2, \sigma_1, \sigma_2) &= v_5(-L/2-D; E_2), \\ \frac{\mathrm{d}}{\mathrm{d}\xi} v_2(-L/2-D; B_1, B_2, \sigma_1, \sigma_2) &= \frac{\mathrm{d}}{\mathrm{d}\xi} v_5(-L/2-D; E_2), \\ v_2(-L/2; B_1, B_2, \sigma_1, \sigma_2) &= v_3(-L/2; C_{21}, C_{22}), \\ \frac{\mathrm{d}}{\mathrm{d}\xi} v_2(-L/2; B_1, B_2, \sigma_1, \sigma_2) &= \frac{\mathrm{d}}{\mathrm{d}\xi} v_3(-L/2; C_{21}, C_{22}), \\ v_1(L/2; A_1, A_2, \delta_1, \delta_2) &= v_3(L/2; C_{21}, C_{22}), \\ \frac{\mathrm{d}}{\mathrm{d}\xi} v_1(L/2; A_1, A_2, \delta_1, \delta_2) &= \frac{\mathrm{d}}{\mathrm{d}\xi} v_3(L/2; C_{21}, C_{22}), \\ v_1(L/2+D; A_1, A_2, \delta_1, \delta_2) &= v_4(L/2+D; D_2), \\ \frac{\mathrm{d}}{\mathrm{d}\xi} v_1(L/2+D; A_1, A_2, \delta_1, \delta_2) &= \frac{\mathrm{d}}{\mathrm{d}\xi} v_4(L/2+D; D_2), \end{aligned} \quad (14)$$

and the normalization condition requires that

$$\int_{-\infty}^{+\infty} \left( |u(\xi)|^2 + |v(\xi)|^2 \right) \mathrm{d}\xi = 1. \quad (15)$$

In Eq. (12), there are seventeen parameters $A_1$, $A_2$, $\delta_1$, $\delta_2$, $B_1$, $B_2$, $\sigma_1$, $\sigma_2$, $C_{11}$, $C_{12}$, $C_{21}$, $C_{22}$, $D_1$, $D_2$, $E_1$, $E_2$ and the propagation constant $\beta$, which can be calculated by solving numerically Eqs. (13), (14), and (15). Once these parameters are determined, one can obtain the exact optical modes for the double-channel waveguide with PT symmetry.

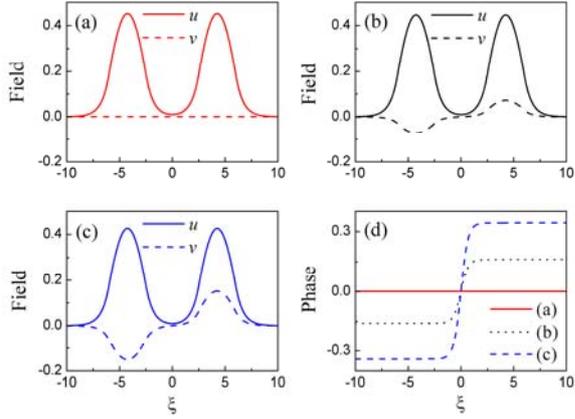
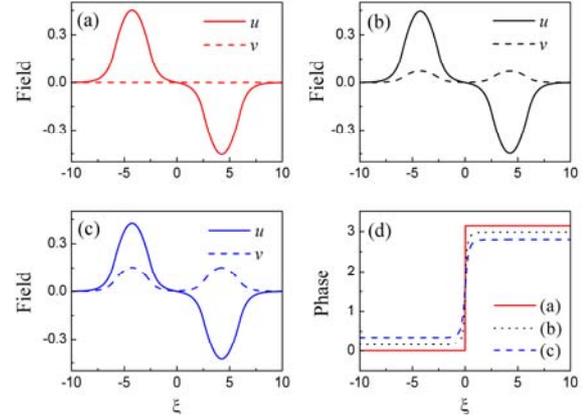

Fig. 2 – The distribution plots of the real and the imaginary parts of optical mode with symmetric real part for: a) $W_0=0$; b) $W_0=0.00005$; c) $W_0=0.0001$, respectively. Note that the corresponding propagation constants are $\beta=2.07251$, $2.07250$, and $2.07247$, respectively; d) the corresponding phases. Here the system parameters are $D=3.5$, $L=5$ and $U_0=2.5$.

Fig. – 3 The distribution plots of the real and imaginary parts of optical mode with antisymmetric real part for: a) $W_0=0$; b) $W_0=0.00005$; c) $W_0=0.0001$, respectively, where the corresponding propagation constant is $\beta=2.07221$, $2.07221$, and $2.07224$, respectively; d) the corresponding phases. Here the system parameters are the same as in Fig. 2.

Figures 2 and 3 present the distribution plots of two fundamental optical modes with symmetric and antisymmetric real parts for the different gain/loss parameter $W_0$, respectively. It should be pointed out that the two modes are reduced from the symmetric and the antisymmetric optical modes in Ref. [21], respectively, as shown in Figs. 2a and 3a. From them, it can be seen that, with the increasing of the gain/loss parameter $W_0$, for the mode with symmetric real part, the real part $u$ always remains a symmetric function and the imaginary part $v$ is antisymmetrically increasing, as shown in Figs. 2b and 2c, whereas, for the mode with antisymmetric real part, the real part $u$ always remains an antisymmetric function and the imaginary part $v$ is symmetrically increasing, as shown in Figs. 3b and 3c. In addition, we also find that the phase difference at the central position $\xi=0$ increases for the mode with symmetric real part, while the phase difference decreases for the mode with antisymmetric real part, as shown in Figs. 2d and 3d, respectively.

The propagation constants of the two modes with symmetric and antisymmetric real parts as a function of the gain/loss parameter $W_0$ are plotted in Fig. 4a. From it, one can see that, with the increasing of the gain/loss parameter, the propagation constants of the two modes close towards each other, and eventually merge together at a threshold value $W_0^{th}=0.00016$ (for our choice of the parameters). Below the threshold value the propagation constants of the two modes are real, however, once the gain/loss parameter exceeds the threshold value they become complex. Thus, the threshold value corresponds to a branch point, i. e., an exceptional point, at which *PT symmetry is spontaneously broken*. Because the two modes at the branch point have the same propagation constant, they form a pair of degenerate modes, as shown in Figs. 4b and 4c. By comparing Fig. 4b and Fig. 4c, one find that the two degenerate modes have only the phase difference of $\pi/2$. Note that this degenerate mode completely differ from the degenerate state in the nonlinear case [22].

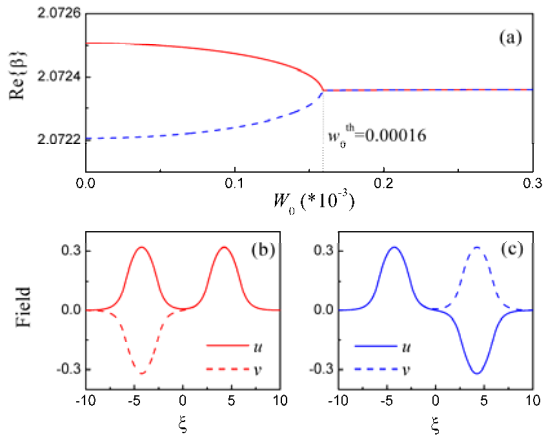

Fig. 4 – a) The real part of propagation constant $\beta$ *versus* the gain/loss parameter $W_0$; b) and c) the distribution plots of optical modes shown in Fig. 2 and Fig. 4 for $W_0=W_0^{th}$, respectively. Here the system parameters are the same as in Fig. 2.

Furthermore, the excited optical modes in the double-channel waveguide structure are discussed. Figures 5 and 6 show the distribution plots of two excited optical modes with symmetric and antisymmetric real parts for different gain/loss parameters $W_0$, respectively. Similarly, one can see that the real part $u$ always keeps a symmetric profile and the imaginary part $v$ is an antisymmetric function with the increasing of the gain/loss parameter $W_0$, for the excited mode with symmetric real part, as shown in Figs. 5b and 5c. For the excited mode with antisymmetric real part, the real part $u$ always keeps an antisymmetric profile and the imaginary part $v$ is a symmetric function with the increasing of the gain/loss parameter $W_0$, as shown in Figs. 6b and 6c. The corresponding phase distributions exhibit a more complex structure, as shown in Figs. 5d and 6d, respectively. From the results presented in Figs. 5 and 6, one can see that the phase difference at the central position $\xi = 0$ increases for the mode with symmetric real part, while decreases for the mode with antisymmetric real part. However, the phase differences at the nodes keep invariance.

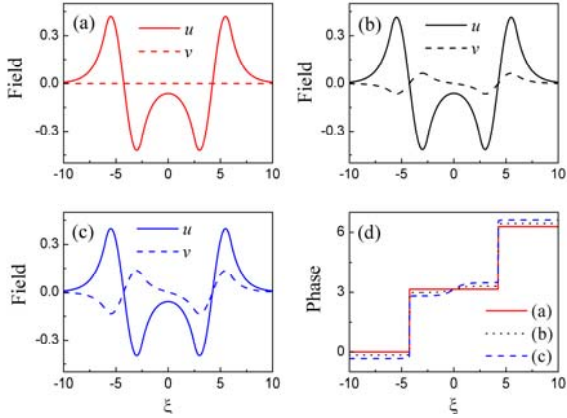

Fig. 5 – The distributions of excited mode with a symmetric real part for: a) $W_0=0$; b) $W_0=0.002$; c) $W_0=0.004$, where the corresponding propagation constant is $\beta=0.9031$, $0.90278$, and $0.90157$, respectively; d) the corresponding phases. Here the system parameters are the same as in Fig. 2.

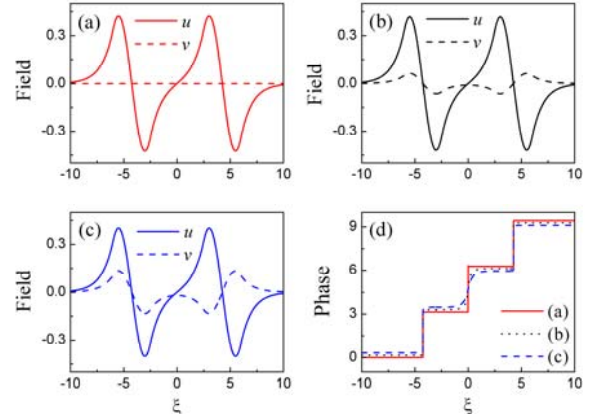

Fig. 6 – The distributions of excited mode with antisymmetric real part for: a) $W_0=0$; b) $W_0=0.002$; c) $W_0=0.004$, where the corresponding propagation constant is $\beta=0.89554$, $0.89586$, and $0.89707$, respectively; d) the corresponding phases. Here the system parameters are the same as in Fig. 2.

For the two excited optical modes with symmetric and antisymmetric real parts, the dependence of their propagation constants on the gain/loss parameter is also studied, and the relevant results are summarized in Fig. 7. The results show that there exists a *threshold value* $W_0^{th} = 0.005$ (for our choice of the parameters). Below the threshold value the propagation constants of the two excited modes are real, but when the gain/loss parameter exceeds the threshold value they become complex. Thus a *branch point* appears at the threshold value, as shown in Fig. 7a. The two excited modes at the branch point have the same propagation constant, so they present a pair of degenerate excited modes, as shown in Figs. 7b and 7c. It is similarly found that the two degenerate excited modes have only the phase difference of $\pi/2$, which also differ from the degenerate excited mode in the nonlinear case [22]. It should be pointed out that the threshold value for the excited optical modes is larger than that of the fundamental optical modes, which means that in the double-channel waveguide with PT symmetry, it is possible that the fundamental mode does not exist, but the excited mode may exist.

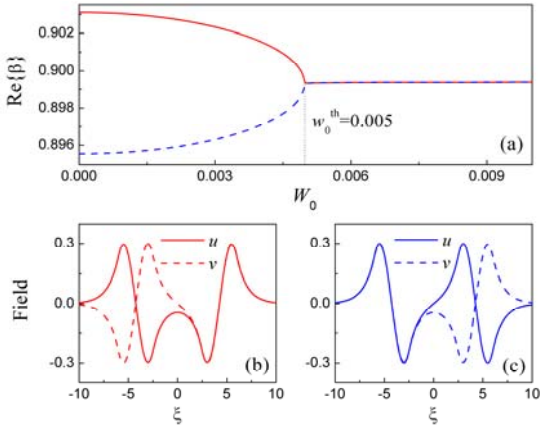

Fig. 7 – a) The real part of propagation constant $\beta$ versus the gain/loss parameter $W_0$; b) and c) the distribution plots of optical modes shown in Fig. 7 and in Fig. 9 for the threshold value $W_0=W_0^{th}$, respectively. The system parameters are the same as in Fig. 2.

It should be pointed out that all modes shown in Figs. 2–7 are symmetry-preserving modes because when $W_0 = 0$ they can be reduced to what was previously shown in Ref. [21]. In order to better understand the properties of the optical modes in the double-channel waveguides with PT symmetry, we also discuss the influences of the system parameters on the optical modes.

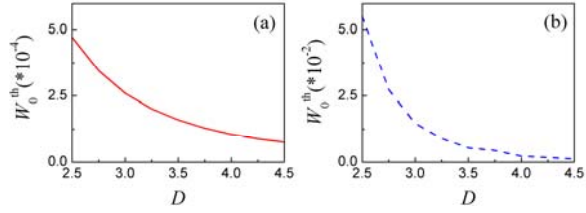

Fig. 8 – The dependence of the threshold value of: a) the fundamental modes; b) the excited modes on the channel-width $D$, respectively. Here the system parameters are $L=5$ and $U_0=2.5$.

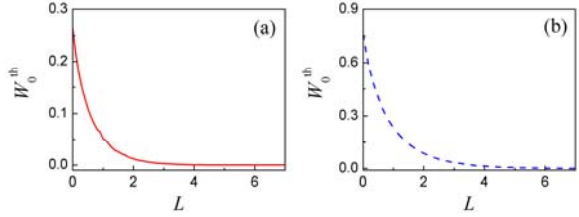

Fig. 9 – The dependence of the threshold value of: a) the fundamental modes; b) the excited modes on the channel separation $L$, respectively. Here the system parameters are $D=3.5$ and $U_0=2.5$.

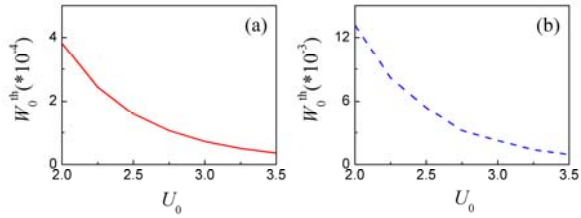

Fig. 10 – The dependence of the threshold value of: a) the fundamental modes; b) the excited modes on the modulation depth $U_0$, respectively. Here the system parameters are $L=5$ and $D=3.5$.

Figures 8–10 present the dependence of the threshold value for both the fundamental and the excited modes, on the channel-width $D$, the channel separation $L$ and the modulation depth $U_0$, respectively. From these figures, one find that the value of $W_0^{th}$ varies strongly with the change of $D$, $L$, and $U_0$. This means that one can extend the region of the gain/loss parameter $W_0$ by tuning properly the system's parameters. For example, we can enhance the threshold value of the gain/loss parameter by decreasing the separation between the channels, as shown in Fig. 9.

In the following, we discuss the beam dynamics. For the linear system, the superposition of two eigenmodes with symmetric and antisymmetric real parts is also a solution of the system. We assume that the solution with symmetric real part is of the form $(u_1+iv_1)\exp(i\beta_1 z)$, and the solution with antisymmetric real part is of the form $(u_2+iv_2)\exp(i\beta_2 z)$. Note that the two modes have the following symmetry properties on the axis $\xi=0$, i. e., $u_1 \to u_1$, $v_1 \to -v_1$ and $u_2 \to -u_2$, $v_2 \to v_2$. We consider both the sum mode and the difference mode for the above two modes as follows

$$\psi_+ = \frac{1}{\sqrt{2}}[(u_1+iv_1)e^{i\beta_1 z}+(u_2+iv_2)e^{i\beta_2 z}], \tag{16}$$

$$\psi_- = \frac{1}{\sqrt{2}}[(u_1+iv_1)e^{i\beta_1 z}-(u_2+iv_2)e^{i\beta_2 z}], \tag{17}$$

in which the corresponding power distribution for the sum mode and the difference mode are

$$|\psi_+|^2 = \frac{1}{2}(u_1^2+v_1^2+u_2^2+v_2^2)+(u_1 u_2+v_1 v_2)\cos(\Delta\beta z)+(u_1 v_2-v_1 u_2)\sin(\Delta\beta z), \tag{18}$$

$$|\psi_-|^2 = \frac{1}{2}(u_1^2 + v_1^2 + u_2^2 + v_2^2) - (u_1 u_2 + v_1 v_2)\cos(\Delta\beta z) - (u_1 v_2 - v_1 u_2)\sin(\Delta\beta z), \quad (19)$$

respectively, where $\Delta\beta = \beta_1 - \beta_2$. The case $W_0=0$ corresponds to the conventional Hermitian system, i. e., $v_1 = v_2 = 0$. Thus, Eqs. (18) and (19) can be reduced to

$$|\psi_+|^2 = \frac{1}{2}(u_1^2 + u_2^2) + u_1 u_2 \cos(\Delta\beta z), \quad (20)$$

$$|\psi_-|^2 = \frac{1}{2}(u_1^2 + u_2^2) - u_1 u_2 \cos(\Delta\beta z), \quad (21)$$

respectively.

From the expressions (20) and (21), one can see that the wave propagation exhibits *left-right symmetric oscillations* with a beat length of $L = 2\pi/\Delta\beta$, which means a reciprocal wave propagation, as shown in Figs. 11a and 11b for the superpositions of the two fundamental state modes, and in Figs. 12a and 12b for the superpositions of the two excited state modes. In this case, the total power is conserved.

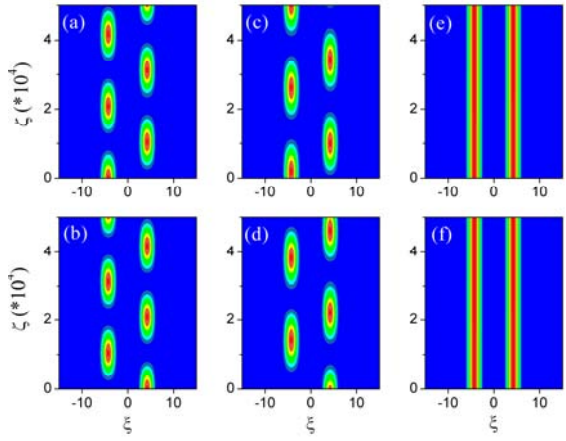
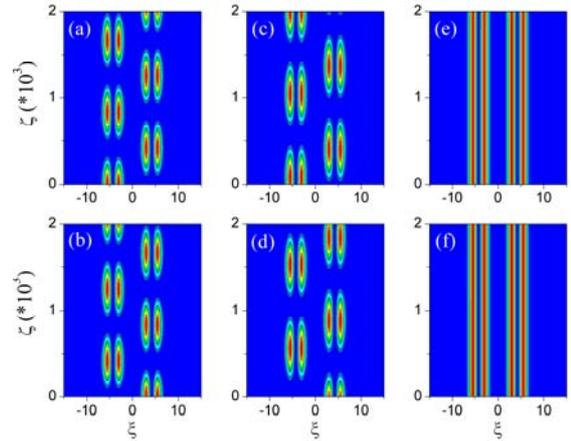

Fig. 11 – The evolution plots of the power distributions $|\psi_+|^2$ (top panel) and $|\psi_-|^2$ (bottom panel) of the superposition of two fundamental state modes for the different gain/loss parameter $W_0$, where $W_0 = 0$ in a) and b) $W_0 = 0.00008$, in c) and d), and $W_0 = 0.00016$ in e) and f). Here the system parameters are $L = 5$, $D = 3.5$ and $U_0 = 2.5$.

Fig. 12 – The evolution plots of the power distributions $|\psi_+|^2$ (top panel) and $|\psi_-|^2$ (bottom panel) of the superposition of two excited state modes for the different gain/loss parameter $W_0$, where $W_0 = 0$ in a) and b) $W_0 = 0.0025$ in c) and d), and $W_0 = 0.005$ in e) and f). Here the system parameters are $L = 5$, $D = 3.5$ and $U_0 = 2.5$.

With the increasing of the gain/loss parameter $W_0$, as expressed by Eqs. (18) and (19), the beat length becomes larger because the propagation constant difference $\Delta\beta$ decreases (see the plots of the propagation constant versus the gain/loss parameter in Figs. 4a and 7a), and the wave propagation exhibits the characteristic feature of the *nonreciprocal propagation*, as shown in Figs. 11c and 11d for the superposition of the fundamental state modes, and in Figs. 12c and 12d for the superposition of the excited state modes [7]. In this case, the total power is no longer conserved.

However, when $W_0$ is increased to the threshold value $W_0^{th}$, the beat length is infinite because the two propagation constants at the threshold value are the same. In this case, the power in the double-channel waveguides shares the same distribution because of the superposition of the two degenerate modes (see Figs. 4b, 4c and Figs. 7b, 7c), in which the corresponding power distributions of $|\psi_+|^2$ and $|\psi_-|^2$ are of the form

$$|\psi_+|^2 = \frac{1}{2}\left[(u_1 + u_2)^2 + (v_1 + v_2)^2\right], \quad |\psi_-|^2 = \frac{1}{2}\left[(u_1 - u_2)^2 + (v_1 - v_2)^2\right],$$

respectively. Thus, *bound states* are formed in the double-channel waveguides, as shown in Figs. 11e and 11f for the fundamental state modes, and Figs. 12e and 12f for the excited state modes.

## 4. CONCLUSIONS

In summary, we have analytically discussed the fundamental and the excited optical modes in a double-channel waveguide with PT symmetry. The results have shown that these modes are symmetry-preserving solutions. The spontaneous PT symmetry breaking point (the occurrence of a threshold value for the gain/loss parameter) has been found by employing the dependence of the propagation constant on the gain/loss parameter. Also, the dependence of the threshold value on the structure parameters has been studied, and it was found that the threshold value for the excited mode is larger than that of the fundamental mode. The obtained results have shown that the PT symmetry condition may result in a large gain/loss effect by properly tuning the waveguide structure parameters. Moreover, the beam dynamics has been investigated in detail.

## ACKNOWLEDGEMENT

This research was supported by the National Natural Science Foundation of China grant 61078079, the Shanxi Scholarship Council of China Grant No. 2011-010, and the Provincial Program of Undergraduate Innovative Training of Shanxi University.